\documentclass{article}

\usepackage{amssymb,amsmath,xcolor,graphicx,xspace,colortbl,ragged2e,rotating} %
\usepackage{amsmath}  
\usepackage{amssymb,amsmath,xcolor,graphicx,xspace,colortbl,ragged2e,rotating}  
\usepackage{color}  
\usepackage{tabulary}  
\graphicspath{{october_11 2019_graphics/}{october_11 2019_tcache/}{october_11 2019_gcache/}}
\DeclareGraphicsExtensions{.pdf,.eps,.ps,.png,.jpg,.jpeg}
\graphicspath{{july_24 2019_graphics/}{july_24 2019_tcache/}{july_24 2019_gcache/}}
\DeclareGraphicsExtensions{.pdf,.eps,.ps,.png,.jpg,.jpeg}
\graphicspath{{may_16 19_graphics/}{may_16 19_tcache/}{may_16 19_gcache/}}
\DeclareGraphicsExtensions{.pdf,.eps,.ps,.png,.jpg,.jpeg}

\begin{document}
\title{A relational time-symmetric framework for analyzing the quantum computational speedup}
\author{G. Castagnoli\protect\footnote{
Elsag Bailey ICT Division and Quantum Information Laboratory Via Puccini 2, 16154 Genova, Italy}, E. Cohen\protect\footnote{
Faculty of Engineering and the Institute of Nanotechnology and Advanced Materials, Bar Ilan University, Ramat Gan 5290002, Israel},
A. K. Ekert\protect\footnote{  Mathematical Institute, University of Oxford
and Centre for Quantum Technologies, National University of Singapore}, A. C. Elitzur\protect\footnote{
Institute for Quantum Studies, Chapman University, Orange, CA 92866, USA and Iyar, The Israeli Institute for Advanced Research, POB 651, Zichron, Ya'akov
3095303, Israel}}
\maketitle
\begin{abstract}The usual representation of quantum algorithms is limited to the process of solving the problem. We extend it to the process of setting the problem.
Bob, the problem setter, selects a problem-setting by the initial measurement. Alice, the problem solver, unitarily computes the corresponding solution
and reads it by the final measurement. This simple extension creates a new perspective from which to see the quantum algorithm. First, it highlights the relevance of
time-symmetric quantum mechanics to quantum computation: the problem-setting and problem solution, in their quantum version, constitute pre- and post-selection,
hence the process as a whole is bound to be affected by both boundary conditions. Second, it forces us to enter into relational quantum mechanics. There
must be a representation of the quantum algorithm with respect to Bob, and another one with respect to Alice, from whom the outcome of the initial measurement,
specifying the setting and thus the solution of the problem, must be concealed. Time-symmetrizing the quantum algorithm to take into account both boundary
conditions leaves the representation to Bob unaltered. It shows that the representation to Alice is a sum over histories in each of which she remains shielded from the information coming
to her from the initial measurement, not from that coming to her backwards in time from the final measurement. In retrospect, all is as if she knew in advance, before performing
her problem-solving action, half of the information that specifies the solution of the problem she will read in the future and could use this information
to reach the solution with fewer computation steps (oracle queries). This elucidates the quantum computational speedup in all the quantum algorithms examined.
\end{abstract}

\section{Introduction}
The term \textit{quantum computational speedup} accounts for the fact that solving certain problems can be more efficient quantumly
than classically. For example, Bob, the problem setter, hides a ball in a chest of four drawers. Alice, the problem solver, is to locate it by opening drawers
(by \textit{querying the oracle}: is the ball in that drawer?). In the classical case, Alice may need to open up to three drawers: if the ball
has not yet been found, it must be in the fourth. The quantum algorithm devised by Grover $\left [1\right ]$ always requires opening just one drawer. It is therefore said to
yield a quantum computational speedup, meaning that when encoding the classical problem with quantum states and using quantum operations to solve it, the
number of drawers that need to be opened is reduced. 

The four drawer problem is an example of the \textit{oracle problem}s.
In the most general case, Bob selects a function from a set of functions also known to Alice and provides Alice with the black box (oracle) that, given
an argument, computes the value of the function. Alice, who does not know Bob's choice, is to find a characteristic of the function by computing the value of the function for different values of the argument (i.e., by performing \textit{oracle queries}). In Grover's problem, the set of functions is constituted by the Kronecker
functions $\delta  \left (b ,a\right )$ for all the possible values of $b$, where $b$ is the number of the drawer with the ball -- the characteristic of the function to be found by Alice -- and $a$ that of the drawer opened by her. Bob selects a value of $b$ and gives Alice the black box that computes the corresponding Kronecker function. Alice is to identify $b$ by computing $\delta  \left (b ,a\right )$ for different values of $a$ (i. e. by opening different drawers). The present work is an analysis of the speedups achieved in the quantum solution of oracle
problems. 

All quantum algorithms devised so far have been found by means of ingenuity. Although some important unifications have been
a posteriori identified, e.g. $\left [2\right ]$, no general one -- holding for any quantum algorithm -- has been
discovered until now. In particular, given a generic oracle problem, it would be desirable to have a method for estimating the number of oracle queries
required to solve it in an optimal quantum way. This is the so-called quantum query complexity problem, see e.g. $\left [3\right ]$, and is still open. 

In mainstream literature,
there are various research areas concerning quantum computation. The main one has been of course the search for new quantum algorithms. For what concerns
the study of the quantum speedup, we shall mention two approaches that share our present objectives: 

1) Looking for a fundamental
explanation of the speedup, in particular for possible relationships between it and other fundamental quantum features, such as entanglement -- $\left [4 -6\right ]$. This is still work in progress, as so far no complete relationship
between entanglement (or discord) and speedup has been proven in full generality $\left [6\right ]$. 

2)\ Looking
for lower and/or upper bounds to the quantum computational complexity of classes of problems. This is the approach of quantum computational complexity theory,
see e.g. $[3 ,7]$. The classes found so far are still far away from covering all the oracle problems.
Moreover, since their identification is entirely analytic, the approach in question does not seem to shed much light upon the physical reason for the speedup.

The present work immerses a previous retrocausal interpretation of the speedup $\left [8 -10\right ]$ into the context of time-symmetric quantum mechanics. In particular,
this allows to analyze its physical viability. Thus, we too shall pursue a fundamental analysis of the speedup, but in a new representation of quantum algorithms,
as follows. 

The usual representation is limited to the process of computing the solution of the problem. Let us refer to the drawers
and ball problem. Alice (the problem solver) works with a quantum register $A$ meant to contain $a$, the number of the drawer that she wants to open (i. e. the number to query the oracle with), and eventually the solution of the problem
(the number of the drawer with the ball $b$). In its initial state, register $A$ contains an arbitrary drawer number. The process of setting the problem (of selecting the value of $b$) and the corresponding initial measurement are missing. 

To complete the representation, we add a quantum
register $B$, under the control of Bob (the problem setter), meant to contain the number of the drawer with the ball $b$ (the \textit{problem-setting}). Initially, the state of register $B$ is a superposition (or indifferently a mixture, see further on) of all the possible numbers of the drawer with the ball (the
reason of this choice will soon become clear). By an initial measurement of the content of register $B$, Bob selects a number of the drawer with the ball at random. He could then unitarily change it into a desired number, but for
the sake of simplicity we omit this operation. 

Alice is forbidden to measure the content of register $B$, it would tell her the problem-setting and thus also the corresponding solution of the problem (here both the number of the drawer with the
ball) without the need of performing oracle queries. She can access the information contained in $B$ only by oracle queries (by opening drawers). Let us think of a generic number of drawers; each query acquires some of the information
about the number of the drawer with the ball contained in register $B$ and puts it in register $A$. Eventually also this register will contain the number of the drawer with the ball -- namely the solution of the problem. Alice acquires it by measuring the content of $A$. With probability one of reading the solution, there is a unitary transformation between the initial and final measurement outcomes.
The process between them is physically reversible since no information is destroyed along it. 

Extending the representation has two
consequences: 

I) The extended representation works for Bob and any external observer, not for Alice. It would tell her the number
of the drawer with the ball selected by Bob before she begins to compute it. In computer science language, to Alice, the number of the drawer with the ball
selected by Bob must be hidden inside the black box. 

II) Highlighting the relevance of time-symmetric quantum mechanics to quantum
computation: the problem-setting and problem solution, in their quantum version, constitute pre- and post-selection, hence the process between them is bound
to be affected by both boundary conditions. Since the process in question is reversible in time, for reasons of time-symmetry it should be evenly affected.

Point (I) is addressed as follows. We conceal the number of the drawer with the ball from Alice by postponing the projection of the
quantum state due to initial Bob's measurement to the end of the unitary part of her problem-solving action -- we use a well-known degree of freedom of
the quantum description $\left [11\right ]$.
In this way, to Alice, the state of register $B$ immediately after the initial measurement remains a superposition of all the possible numbers of the drawer with the ball, namely
a state of complete ignorance of the number selected by Bob. Correspondingly, the unitary part of Alice's action $\hat{\mathbf{U}}$ generates a superposition of quantum algorithms, one for each possible number of the drawer with the ball. Eventually the final Alice's measurement
selects the solution corresponding to the number already selected by Bob. 

Point (II) is addressed as follows. 

What is determined by either the initial or the final measurement is the number of the drawer with the ball (selected out of $2^{n}$ possible drawer numbers), which is here the outcome of both the initial and final measurements. The unitary
transformation $\hat{\mathbf{U}}$ represents the \textit{quantum computational network} and is of course fixed. 

In the usual way of
thinking, the selection of the $n$ bits that specify the number of the drawer with the ball is all ascribed to the initial measurement; the measurement outcome
then propagates forward in time by $\hat{\mathbf{U}}$ until becoming the outcome of the final measurement -- the solution of the problem (that same number but in register $A$). In the present context, this way of thinking has two shortcomings: (i) it introduces a preferred direction of time and causality
in a reversible physical process and (ii) it takes into account only one of the two boundary conditions that should be taken into account according to time-symmetric
quantum mechanics. 

The way out is to time-symmetrize the process in such a way that the initial and final measurements evenly contribute
to its determination. We should ascribe the selection of half of the $n$ bits that specify the number of the drawer with the ball to the initial measurement and the other half to the final measurement,
in a quantum superposition of all the possible ways of halving. In each of them (in each time-symmetrization instance), the two half selections should be
ordered according to the so called \textit{Parisian zigzag}, namely the logical succession of a forward and a backwards in
time propagation used by Costa de Beauregard $\left [12\right ]$ to explain spatial nonlocality. The outcome of the selection ascribed
to the initial measurement propagates forward in time by $\hat{\mathbf{U}}$ until the right end of it (this is the zig). Here it undergoes the selection ascribed to the final measurement; the outcome of the
latter selection propagates backwards in time by $\hat{\mathbf{U}}^{\dag }$ until the left end of it, namely immediately after the initial measurement (this is the zag). The latter propagation, which inherits
both selections, is an \textit{instance} of the time-symmetrized quantum algorithm. Always being the unitary transformation  $\hat{\mathbf{U}}$, it can be read of course also from left to right, namely as a forward in time propagation, as a time-symmetrization instance of
the quantum algorithm in fact. 

This time-symmetrization procedure: 

i) Leaves unaltered the representation of
the quantum algorithm with respect to Bob and any external observer, which is ordinary in character in the sense that no observer is shielded from any measurement
outcome.

ii) It has consequences in the representation of the quantum algorithm with respect to Alice, the observer of the
final measurement shielded from the outcome of the initial measurement. This representation turns out to be the quantum superposition of all the time-symmetrization
instances, characterized as follows. 

In each instance, just before the beginning of her problem-solving action, Alice remains shielded
from the half information coming to her from the initial measurement, not from the half coming to her back in time from the final measurement. This reduces
the number of elements in the initial superposition of all the possible numbers of the drawer with the ball from the $2^{n}$ of the unsymmetrized algorithm to $2^{n/2}$. In other words, the computational complexity of the problem to be solved by her reduces to locating a ball hidden in $2^{n/2}$ drawers. Since the initial superposition of numbers of the drawer with the ball also represents Alice's ignorance of this number, also
her knowledge of it correspondingly changes. All is as if she knew in advance half of the $n$ bits that specify the number of the drawer with the ball (the half selected by the final measurement) and could use this information
to reach the solution of the problem by opening fewer drawers (by fewer oracle queries). This quantitatively accounts for the speedup of Grover algorithm.
The same mechanism accounts for the speedup of all the quantum algorithms examined, which comprise the major ones and cover both the quadratic and exponential
speedups. 

Note that this time the Parisian zigzag brings us to a form of temporal nonlocality: in each time-symmetrization instance, the selection ascribed to the final
measurement at the end of $\hat{\mathbf{U}}$ nonlocally changes Alice's state of knowledge of the problem-setting at the beginning of  $\hat{\mathbf{U}}$. 

Alice's advanced knowledge faces the problem of reconciling time-reversal symmetry with the principle of causality.
This is a normal problem in time-symmetric quantum mechanics. The present interpretation of the speedup has indeed a precedent in this theory. It has been
inspired by the work of Dolev and Elitzur $\left [13\right ]$ on the non-sequential character of the wave function highlighted
by partial measurement. We conjecture that also the subsequent development of this notion into that of quantum oblivion $\left [14 ,15\right ]$, which is much more general and covers, among other phenomena,
interaction-free measurement and the Aharonov-Bohm effect, is applicable to quantum computation. 

In the following, first we develop
the mathematical framework of the present interpretation of the quantum computational speedup. Then we discuss whether it can be physical. It should be
as physical as the main applications of time-symmetric quantum mechanics. Note that many of these applications have been experimentally verified. Of course,
the (so to speak) empirical verification of the present interpretation of the speedup is that it quantitatively accounts for the speedups of a variety of
quantum algorithms. 

\section{ Mathematical framework }
We review the interpretation of the speedup developed in $\left [8 -10\right ]$ while also introducing some significant clarifications. For simplicity
of presentation, we limit ourselves to the simplest four drawer instance of Grover's algorithm. Generalization to quantum oracle computing is straightforward
and will be later discussed. 

\subsection{ Usual representation}
The usual representation of quantum algorithms is limited to the process of computing the solution of the problem. We consider the four drawer
instance of Grover algorithm. Let us number the drawers $0 ,1 ,2 ,3$, namely $00 ,01 ,10 ,11$ in binary notation. Let the number of the drawer with the ball be $b =01$. The quantum computation of the solution is represented in the following table:
\begin{equation}\begin{array}{ccc}\text{\thinspace \thinspace } & \text{\thinspace \thinspace } & \text{meas.}\hat{A}\text{} \\
\vert 00 \rangle _{A} \left (\vert 0 \rangle _{V} -\vert  1 \rangle _{V}\right ) &  \Rightarrow \hat{U} \Rightarrow  & \vert 01 \rangle _{A} \left (\vert 0 \rangle _{V} -\vert  1 \rangle _{V}\right )\end{array} \label{usual}
\end{equation}

The (quantum) register $A$, under the control of the problem solver Alice, is meant to contain the number of the drawer to query the oracle with [the argument
$a$ of the Kronecker function $\delta  (01 ,a)$] and eventually the solution of the problem -- the number of the drawer with the ball $b$. Its basis vectors are thus $\vert 00 \rangle _{A} ,\vert 01 \rangle _{A} ,\vert 10 \rangle _{A} ,\vert 11 \rangle _{A}$. Register $V$, of basis vectors $\vert 0 \rangle _{V}\text{,}$$\vert 1 \rangle _{V}$, is meant to contain the result of the computation of $\delta  \left (b ,a\right )$ modulo $2$ added to its previous content (note that this is a logically reversible operation that can be implemented by a unitary transformation).
The initial state of the quantum algorithm (on the left) is any sharp state of register $A$ known to Alice (standing for a blank blackboard) tensor product a suitable initial state of register $V$ (see what follows). 

$\hat{U}$ is the unitary part of Alice's problem-solving action. It sends the input state $\vert 00 \rangle _{A}(\vert 0 \rangle _{V} -\vert  1 \rangle _{V})$ into the output state $\vert 01 \rangle _{A}(\vert 0 \rangle _{V} -\vert  1 \rangle _{V})$, with the solution of the problem encoded in register $A$ (follow the horizontal arrows). Although the present analysis of the speedup is independent of it, for the records we provide
the precise form of $\hat{U}$:
\begin{equation*}\hat{U} =\hat{\Im } \hat{F} \hat{H}\text{.}
\end{equation*}

$\hat{H}$ is the \textit{Hadamard transform}; it sends $\vert 00 \rangle _{A}$ into the superposition of all the possible values of $a$, namely into $\vert 00 \rangle _{A} +\vert  01 \rangle _{A} +\vert 10 \rangle _{A} +\vert 11 \rangle _{A}$ -- here and in the following we disregard normalization. 

$\hat{F}$, also a unitary transformation, represents the computation of $\delta  \left (01 ,a\right )$. Being preceded by $\hat{H}$, it is performed in a quantum superposition of all the possible values of $a$. If $a =01\text{}$, namely when the input state of $\hat{F}$ (an element of the superposition) is $\vert 01 \rangle _{A}(\vert 0 \rangle _{V} -\vert  1 \rangle _{V})$, the result of the computation is $1$ that, modulo $2$ added to the previous content of register $V$, changes a previous $0$ into $1$ and vice-versa, thus changes $\vert 01 \rangle _{A}(\vert 0 \rangle _{V} -\vert  1 \rangle _{V})$ into $ -\vert 01 \rangle _{A}(\vert  0 \rangle _{V} -\vert  1 \rangle _{V})$. If $a \neq 01\text{}$, the result of the computation is $0$ and the input state of $\hat{F}$ remains unaltered. From now on we call this operation (namely $\hat{F}$) \textit{function evaluation} rather than \textit{oracle query}. 

Up to this point, the initial state $\vert 00 \rangle _{A}(\vert 0 \rangle _{V} -\vert  1 \rangle _{V})$ has changed into: 

\begin{equation}(\vert 00 \rangle _{A} -\vert  01 \rangle _{A} +\vert 10 \rangle _{A} +\vert  11 \rangle _{A}) \left (\vert 0 \rangle _{V} -\vert  1 \rangle _{V}\right ) \label{buto}
\end{equation}The second term in the superposition of all the possible values of $a$, encoding $a =01$, has changed its sign. If the number of the drawer with the ball had been $b =00$ instead of $b =01$, the first term would have changed its sign, and so on. One can readily see that the four superpositions corresponding to the
four possible numbers of the drawers with the ball ($b =00 ,01 ,10 ,11$) are orthogonal with one another. This means that there is a unitary transformation that changes them into respectively the
(orthogonal ) basis vectors of register $A$, namely $\vert 00 \rangle _{A} ,\vert 01 \rangle _{A} ,\vert 10 \rangle _{A} ,\vert 11 \rangle _{A}$. This is in fact the unitary transformation $\hat{\Im }$ (the so called \textit{inversion about the mean}). In particular $\hat{\Im }$ sends state (\ref{buto}) into the state in the right corner of table (\ref{usual}),
which encodes the solution of the problem.

Eventually Alice acquires the solution by measuring the content of register $A$, namely the observable $\hat{A}$ of eigenstates $\vert 00 \rangle _{A} ,\vert 01 \rangle _{A} ,\vert 10 \rangle _{A} ,\vert 11 \rangle _{A}$ and eigenvalues respectively $00 ,01 ,10 ,11$. The output state $\vert 01 \rangle _{A} \left (\vert 0 \rangle _{V} -\vert  1 \rangle _{V}\right )$, with register $A$ already in an eigenstate of $\hat{A}$, remains unaltered. There is thus a unitary transformation between the initial and final measurement outcomes; the process between
them is reversible as no information is destroyed along it. 

By the way, let us note that the state of register $V$ is always the same and factorized. Therefore we can do without this register. It suffices to keep in mind that the function
evaluation operator $\hat{F}$ sends $\vert a \rangle _{A}$ into $ -\vert a \rangle _{A}$ when $\delta  \left (01 ,a\right )$$ =1$, into itself otherwise. 

Let us compare the present quantum algorithm with its classical correspondent.
Classically, to identify the number of the drawer with the ball, Alice has to open drawers [perform the evaluation of $\delta  \left (01 ,a\right )$] in sequence. In the worst case she has to perform a sequence of three function evaluations. In the quantum case instead, she always
locates the ball with just one function evaluation. The point is that function evaluation is now performed for a quantum superposition of all the possible drawer numbers, as seen above. This is called \textit{quantum parallelism}. It is a feature common to all quantum algorithms. Although it must be an essential reason for the quantum speedup, it does
not explain it quantitatively: given a generic oracle problem, it does not allow to compute the number of function evaluations required to solve it in an
optimal quantum way. 

\subsection{ Extended representation}
We extend the usual representation, limited to the process of solving the problem, to that of setting it. We should add a register $B$, under the control of the problem setter Bob, meant to contain the problem-setting $b\text{}$ (the number of the drawer with the ball). Its basis vectors are thus: $\vert 00 \rangle _{B} ,\vert 01 \rangle _{B} ,\vert 10 \rangle _{B} ,\vert 11 \rangle _{B}$. 

The complete process of setting and solving the problem is: 

\begin{equation}\begin{array}{ccc}\;\text{meas.}\;\hat{B} & \text{\thinspace \thinspace } & \;\text{meas.}\;\hat{A} \\
\left (\vert 00 \rangle _{B} +\vert  01 \rangle _{B} +\vert 10 \rangle _{B} +\vert  11 \rangle _{B}\right )\vert 00 \rangle _{A} & \text{\thinspace \thinspace } & \text{\thinspace \thinspace } \\
\Downarrow  & \text{\thinspace \thinspace } & \text{\thinspace \thinspace } \\
\vert 01 \rangle _{B}\vert  00 \rangle _{A} &  \Rightarrow \hat{\mathbf{U}} \Rightarrow  & \vert 01 \rangle _{B}\vert  01 \rangle _{A}\end{array} \label{ex}
\end{equation}

 Note that we are doing without register $V$, which would remain everywhere in the same factorized state $\left (\vert 0 \rangle _{V} -\vert  1 \rangle _{V}\right )$. The initial state of the quantum algorithm (above the vertical
arrow) is a superposition of all the possible problem problem-settings (numbers of the drawer with the ball) tensor product the sharp state of register
$A$.

We note that the quantum superposition in question could as well be replaced by the mixture of all the possible
numbers of the drawer with the ball; using a superposition simplifies the notation and changes nothing. In fact, under the unitary part of Alice's problem-solving
action, the basis vectors of register $B$ never interfere with one another: function evaluations leave them unaltered and the other unitary transformations do not apply
to register $B$. As a consequence, also the reduced density operator of register $B$ remains unaltered. 

Let the observable $\hat{B}$ be the number contained in register $B$. Its eigenstates and eigenvalues are thus respectively $\vert 00 \rangle _{B} ,\vert 01 \rangle _{B} ,\vert 10 \rangle _{B} ,\vert 11 \rangle _{B}$ and $00 ,01 ,10 ,11$. Bob, the problem setter, measures $\hat{B}$ in the initial state. This projects the initial superposition on an eigenstate of $\hat{B}$ selected at random, say $\vert 01 \rangle _{B}$ (follow the vertical arrow). The corresponding eigenvalue, here $b =01$, is the problem-setting selected by Bob. 

$\hat{\mathbf{U}}$, the unitary part of Alice's problem-solving action, differs from $\hat{U}$ (of the usual representation) only for the fact that function evaluation is now the computation of $\delta  \left (b ,a\right )$. It sends $\vert b \rangle _{B}\vert  a \rangle _{A}$ into $ -\vert b \rangle _{B}\vert  a \rangle _{A}$ when $b =a$, into itself otherwise. In the overall, $\hat{\mathbf{U}}$ sends $\vert 01 \rangle _{B}$$\vert 00 \rangle _{A}$ into $\vert 01 \rangle _{B}\vert  01 \rangle _{A}$, where the state of register $A$ encodes the solution of the problem. Alice acquires it by measuring $\hat{A}$. The output state, $\vert 01 \rangle _{B}\vert  01 \rangle _{A}$, remains unaltered. 

We have already said that the present interpretation of the speedup will be independent of the
form of $\hat{\mathbf{U}}$, the unitary part of Alice's problem-solving action. It will only require that there can be a unitary transformation between the input
and the output. This is always the case since, in the extended representation, the output has a full memory of the input. 

\subsection{ Representation relativized to Alice}
The extended representation of the quantum algorithm works for Bob and any external observer, not for Alice (the problem solver). The state
immediately after the initial measurement, $\vert 01 \rangle _{B}\vert  00 \rangle _{A}$, would tell her the problem-setting (that $b =01$) and thus the solution of the problem before she begins her problem-solving action. To Alice, the problem-setting must be hidden
inside the black box. 

To physically represent the concealment of the problem-setting (the outcome of the initial measurement) from
Alice, we can postpone at the end of her problem-solving action the projection of the quantum state associated with the initial measurement. This is of
course a mathematically legitimate operation provided that the two extremes of the projection undergo the unitary transformation $\hat{\mathbf{U}}$ see $\left [11\right ]$. Alternatively, one can think of postponing the very measurement
of $\hat{B}$ to the time of the measurement of $\hat{A}$; we should keep in mind that the reduced density operator of register $B$ remains unaltered along $\hat{\mathbf{U}}$  and that $\hat{A}$ and $\hat{B}$ commute. In any way, the representation with respect to Alice becomes:

\begin{equation}\begin{array}{ccc}\text{meas.}\;\hat{B} & \, & \text{meas.}\;\hat{A} \\
\left (\vert 00 \rangle _{B} +\vert  01 \rangle _{B} +\vert 10 \rangle _{B} +\vert  11 \rangle _{B}\right )\vert 00 \rangle _{A} &  \Rightarrow \hat{\mathbf{U}} \Rightarrow  & \vert 00 \rangle _{B}\vert  00 \rangle _{A} +\vert 01 \rangle _{B}\vert  01 \rangle _{A} +\vert 10 \rangle _{B}\vert  10 \rangle _{A} +\vert 11 \rangle _{B}\vert  11 \rangle _{A} \\
\, & \, & \Downarrow  \\
\, & \, & \vert 01 \rangle _{B}\vert  01 \rangle _{A}\end{array}
\end{equation}
The input state of $\hat{\mathbf{U}}$ (the unitary part of Alice's problem-solving action) remains a state of maximal indetermination of the problem-setting. It represents
Alice's complete ignorance of the number of the drawer with the ball selected by Bob. Under $\hat{\mathbf{U}}$ (horizontal arrows), this input state evolves into the quantum superposition of four tensor products, each the product of a number
of the drawer with the ball and the corresponding solution (that same number but in register $A$). Eventually, the final measurement of $\hat{A}$ projects this superposition on $\vert 01 \rangle _{B}\vert  01 \rangle _{A}$, the tensor product of the number of the drawer with the ball already selected by Bob and the corresponding solution (vertical arrow).
Note that this projection is indifferently that due to the initial Bob's measurement, postponed. We will come back to this point further on. 

\subsection{ Time symmetrization}
With probability one of reading the solution, there is a unitary transformation ($\hat{\mathbf{U}}$) between the initial and final measurement outcomes. In other words the pre-selected and post-selected quantum evolutions coincide
with one another. 

In the customary way of thinking (we call it \textit{alternative \#1}), the information that specifies
the problem-setting and thus the corresponding solution is all selected by the initial measurement of $\hat{B}$. The measurement outcome then propagates forward in time by $\hat{\mathbf{U}}$ until becoming the outcome of the final measurement of  $\hat{A}$. Correspondingly, the projection
of the quantum state is all ascribed to the initial measurement (as in table \ref{ex}); the final measurement would
select/project nothing. Note that Occam's razor is also implicit in this way of thinking -- the possibility that the same information is selected twice,
by both the initial and final measurements, is excluded. 

Alternative \#2 is the one symmetric in time with respect to the first. One
should assume that the information is all selected by the final measurement. Then the measurement outcome propagates backwards in time by $\hat{\mathbf{U}}^{\dag }$ until becoming the outcome of the initial measurement. 

Note that the observer of the initial measurement would be unable to distinguish between alternatives \#1 and \#2. In any case he sees the same measurement outcome, $\vert 01 \rangle _{B}\vert 00 \rangle _{A}$, whether it has been selected by the initial measurement or is the backwards in time propagation of a selection performed by the final measurement he cannot tell. However, since the process between the initial and final measurement outcomes is reversible, in principle either alternative would have the defect of introducing a preferred direction of time.

Alternative \#3 is the one in accordance with time-symmetric quantum mechanics $\left [13 -25\right ]$. The problem-setting and problem solution, in their quantum version
respectively the outcomes of the initial and final measurements, constitute pre- and post-selection, hence the process between them is bound to be affected
by both boundary conditions (by both the initial and final measurements). In the present case, it should be affected in an even way for reasons of symmetry, since
the process between the initial and final measurement outcomes is reversible in time. 

Note that $\hat{\mathbf{U}}$, which represents the quantum computational network, is fixed; what is determined by the initial or the final measurement is the number
of the drawer with the ball. We should therefore assume that half of the information that specifies this number is selected by
the initial measurement and propagates forward in time and the other half by the final measurement and propagates backwards in time, in a quantum superposition
of all the possible ways of halving. This is the key assumption of the present interpretation of the speedup. 

From an operational
standpoint, we should assume that the initial measurement of $\hat{B}$ and the final measurement of $\hat{A}$, in presence of each other (contextually), reduce to partial measurements that evenly and non-redundantly contribute to the selection
of the information. For example, the initial measurement of $\hat{B}$ could reduce to that of $\hat{B}_{l}$ (the left digit of the number contained in register $B$) and the final measurement of $\hat{A}$ to that of $\hat{A_{r}}$ (the right digit of the number contained in register $A$). Or vice-versa, etc. 

The two corresponding propagations should be ordered according to the so called
Parisian zigzag $\left [12\right ]$. First, we should measure $\hat{B}_{l}$ in the initial state and propagate the measurement outcome forward in time by $\hat{\mathbf{U}}$. Then we should measure $\hat{A_{r}}$ in the output of state of $\hat{\mathbf{U}}$ and propagate the measurement outcome backwards in time by $\hat{\mathbf{U}}^{\dag }$. The latter propagation, which inherits the selections performed by both partial measurements, is an \textit{instance} of
the time-symmetrized quantum algorithm. Of course it can also be read as a forward in time propagation, in fact by $\hat{\mathbf{U}}$. Eventually, we should take a uniform superposition of all the possible instances, ways of evenly sharing the selection of the problem-setting
and the corresponding solution. 

This time-symmetrization procedure leaves the quantum algorithm with respect to Bob, who is not shielded
from any measurement outcome, unaltered. It shows the the quantum algorithm with respect to Alice, who is shielded from the outcome of the initial measurement,
is a superposition of time-symmetrization instances in each of which the computational complexity of the problem to be solved by her is reduced, as follows.

\subsection{ The Parisian zigzag}
Let the problem-setting selected by Bob be $b =01$. We consider the case that the initial measurement of $\hat{B}$ reduces to that of $\hat{B}_{l}$ and the final measurement of $\hat{A}$ to that of $\hat{A_{r}}$. The time symmetrization of the quantum algorithm to Alice is then given by the following zigzag diagram:

\begin{equation}\begin{array}{ccc}\;\text{meas. of}\;\hat{B}_{l} & \text{\thinspace \thinspace } & \;\text{meas. of}\;\hat{A_{r}} \\
\left (\vert 00 \rangle _{B} +\vert  01 \rangle _{B} +\vert 10 \rangle _{B} +\vert  11 \rangle _{B}\right )\vert 00 \rangle _{A} &  \Rightarrow \hat{\mathbf{U}} \Rightarrow  & \vert 00 \rangle _{B}\vert  00 \rangle _{A} +\vert 01 \rangle _{B}\vert  01 \rangle _{A} +\vert 10 \rangle _{B}\vert  10 \rangle _{A} +\vert 11 \rangle _{B}\vert  11 \rangle _{A} \\
\text{\thinspace \thinspace } & \text{\thinspace \thinspace } & \Downarrow  \\
\left (\vert 01 \rangle _{B} +\vert  11 \rangle _{B}\right )\vert 00 \rangle _{A} &  \Leftarrow \hat{\mathbf{U}}^{\dag } \Leftarrow  & \vert 01 \rangle _{B}\vert  01 \rangle _{A} +\vert 11 \rangle _{B}\vert  11 \rangle _{A}\end{array} \label{sya}
\end{equation}The projection of the quantum state associated with the initial measurement of $\hat{B}_{l}$ must be postponed at the end of Alice's problem-solving action -- outside table (\ref{sya}) which
is limited to this action. In fact, any information about the problem-setting should be hidden from her (alternatively, we could postpone the very measurement
of $\hat{B}_{l}$). The top line of the diagram is thus the same of table (4). The measurement of $\hat{A_{r}}$ in the output state of $\hat{\mathbf{U}}$, selecting the $1$ of $01$, projects this state on the superposition of the terms ending in $1$ (vertical arrow). Propagating this superposition backwards in time by $\hat{\mathbf{U}}^{\dag }$ (left looking horizontal arrows), yields an instance of the time-symmetrized quantum algorithm to Alice. See the bottom line of table
(\ref{sya}), repeated here for convenience (of course we can replace $ \Leftarrow \hat{\mathbf{U}}^{\dag } \Leftarrow $ by $ \Rightarrow \hat{\mathbf{U}} \Rightarrow $):
\begin{equation}(\vert 01 \rangle _{B} +\vert  11 \rangle _{B})\vert 00 \rangle _{A} \Rightarrow \hat{\mathbf{U}} \Rightarrow \vert 01 \rangle _{B} \vert 01 \rangle _{A} +\vert  11 \rangle _{B}\vert 11 \rangle _{A} . \label{distance}
\end{equation}

By the way, let us say for completeness that the output state $\vert 01 \rangle _{B}\vert  01 \rangle _{A} +\vert 11 \rangle _{B}\vert  11 \rangle _{A}$ will eventually be projected on $\vert 01 \rangle _{B}\vert  01 \rangle _{A}$ by the projection due to the initial measurement of $\hat{B}_{l}$ postponed after the end of Alice's action (outside table \ref{sya}). If not after the end, it would tell her information about the problem setting independently of her action.

For $b =01$, there are in total three time-symmetrization instances, in each of which the problem-setting $01$ pairs with another problem-setting (it is paired with $11$ in the above instance). The superposition of all instances, also for all the possible problem-settings, gives back the original
quantum algorithm to Alice of table (4). In fact the superposition of all the pairs of basis vectors of register
$B$ yields of course the superposition of all the basis vectors. 

\subsection{ Interpretation}
Let us examine the time-symmetrization instance of table (\ref{distance}). It is important to
note that it is also by itself the full time-symmetrization of a quantum algorithm where Bob and Alice actually perform the measurements of respectively
$\hat{B}_{l}$ and $\hat{A_{r}}$. In this case time-symmetrization only consists in propagating forward in time the selection performed by the measurement of $\hat{B}_{l}$ and backwards in time that performed by the measurement of $\hat{A_{r}}$. Of course these two measurements fully determine the problem-setting and the corresponding solution and therefore the quantum algorithm
is still the same in which Bob and Alice (redundantly) measure respectively $\hat{B}$ and $\hat{A}$. Thus, to the end of ascertaining the number of function evaluations required by the quantum algorithm, we can limit ourselves to
considering a single time-symmetrization instance. 

Back to table (\ref{distance}), we can see
that the problem to be solved by the quantum algorithm is now locating the ball hidden in the pair of drawers $\left \{01 ,11\right \}$ -- check the input and output states of $\hat{\mathbf{U}}$. We call this problem, with reduced computational complexity with respect to the original problem, the \textit{reduced
problem}. In equivalent terms, we can say that Alice, the problem solver, ``knows'' that the ball is in the pair of drawers $\left \{01 ,11\right \}$ before beginning her problem-solving action -- Alice's state of knowledge of the problem-setting is now represented by $\vert 01 \rangle _{B} +\vert 11 \rangle _{B}$. She knows one
of the possible halves of the information about the setting/solution of the problem. The computational complexity of the problem
to be solved by her is correspondingly reduced. 

As any other problem, the reduced problem can always be solved quantumly with the
number of function evaluations required to solve it with a reversible classical algorithm\protect\footnote{
We use the term \textit{classical algorithm} instead of just \textit{algorithm}, i.e. \textit{Turing machine}, to avoid confusion
with the term \textit{quantum algorithm}. However, with this, we do not intend to make any comparison between quantum and classical physics. Here
it is quantum physics and classical logic that face each other. }-- just one in the present case. Let us underline the fact
that this simple consideration will allow us to identify the number of function evaluations required to solve any oracle problem in an optimal quantum way without knowing the quantum algorithm (without knowing $\hat{\mathbf{U}}$). 

What we have found until now is an upper bound to the quantum computational complexity of the oracle problem.
In the four drawer case, this upper bound is also a lower one since the problem cannot be solved with less than one function evaluation. 

To
examine the situation more in general, we should consider the generic number of drawers $N =2^{n}$. Evenly sharing between the initial and final measurements the selection of the $n$ bits that specify the number of the drawer with the ball selected by Bob (the basic assumption here) implies that the reduced
problem is locating the ball in $2^{n/2}$ drawers. This requires classically $\ensuremath{\operatorname*{O}} \left (2^{n/2}\right )$ function evaluations. For what we have seen before, this is also an upper bound to the quantum computational complexity of Grover's problem. In other words, the problem of locating the ball in $2^{n}$ drawers can always be solved quantumly with $\ensuremath{\operatorname*{O}} \left (2^{n/2}\right )$ function evaluations.

Now we can ask ourselves whether it could be solved with even fewer function evaluations. The answer must be negative. If the problem solver could solve the original problem with even fewer function evaluations, logically it would be as if she knew in advance more than $n/2\text{}$ of the bits that specify the number of the drawer with the ball. This would go against the starting assumption that the selection of these bits evenly share between the initial and final measurements. 

With the upper and lower bounds coinciding with each other, the number logically required to solve the reduced problem would thus be the number required by the optimal quantum algorithm. By the way, we must make reference to an optimal quantum algorithm since a non optimal one could take any higher number of function evaluations -- see $\left [27 ,28\right ]$. All this is of course in agreement with the fact that Grover algorithm, which itself requires $\ensuremath{\operatorname*{O}} \left (2^{n/2}\right )$ function evaluations, is demonstrably optimal $\left [26 -28\right ]$.

\subsection{ Generalization}
The drawer and ball problem is an example of oracle problem. More in general, Bob chooses a function from a set of functions and gives Alice
the black box that computes it. Alice, who knows the set of functions but not Bob's choice, is to find a characteristic of the function computed by the
black box by performing function evaluations. Most quantum algorithms solve oracle problems.

To compute the the number of function evaluations required to solve the problem in an optimal quantum way, we need to know only the input and output states of the unitary part of the quantum algorithm to Alice. Let $\sigma $ be the set of all the possible problem-settings, $b$, belonging to $\sigma $, the problem-setting selected by Bob, and $s \left (b\right )$, a function of $b$, the solution of the problem. Disregarding normalization, the input and output states of the unitary part of the quantum
algorithm to Alice have always the form: 

\begin{equation*}\vert \psi  \rangle _{I N} =\sum _{b \in \sigma }\vert b \rangle _{B}\vert 0 \ldots  \rangle _{A}\text{,}
\end{equation*}
\begin{equation}\vert \psi  \rangle _{O U T} =\sum _{b \in \sigma }\vert b \rangle _{B}\vert s \left (b\right ) \rangle _{A} . \label{solution}
\end{equation}Here $\vert 0 \ldots  \rangle _{A}$ is an arbitrary initial state of register $A$ the size of the solution of the problem problem $s \left (b\right )\text{}$; it stands for a blank blackboard -- here we chose a sequence of all zeros.

Note that
we can write $\vert \psi  \rangle _{O U T}$ on the basis of the oracle problem alone, it suffices to know the function $s \left (b\right )$; like in the case of Grover's problem, we do not need to know the quantum algorithm that sends $\vert \psi  \rangle _{I N}$ into $\vert \psi  \rangle _{O U T}$. It is enough to know that there can always be a unitary transformation in between since the output has a full memory of the input.

In $\left [10\right ]$, we have shown that the number of function evaluations required by an optimal quantum algorithm can be computed in general as follows:

\textit{We should split
the initial measurement that selects the problem-setting}\textit{ into two partial measurements that evenly and non-redundantly contribute to
the selection of the problem-setting}\textit{ and the corresponding solution. The information (about the problem-setting}\textit{) acquired
by either partial measurement constitutes an instance of Alice's advanced knowledge. The oracle problem can always be solved in an optimal quantum way with the number of function evaluations required by a classical algorithm that benefits from the advanced knowledge in question.} \medskip
\smallskip

From now on we will call this way of computing the number of function evaluations required by an optimal quantum algorithm \textit{the }\textit{advanced knowledge (AK) rule}.

By the way, applying the even
contribution requirement to the selection of both the problem-setting and the solution is for the case that the solution is not a one to one function of
the problem-setting -- the transformation can be unitary also in this case $\left [10\right ]$.

Applied to equations (\ref{solution}),
the AK rule yields the number of function evaluations in question. Note that the projection of the quantum state
associated with each partial measurement has the only effect of replacing $\sigma $ by a $\sigma ^{ \prime } \subset \sigma $ in both $\vert \psi  \rangle _{I N}$ and $\vert \psi  \rangle _{O U T}$. Therefore, equations (\ref{solution}) tell us the contribution of any partial measurement to
the determination of both the problem-setting and the corresponding solution. Summing up, the two partial measurement we are looking for can be identified
by working only on $\vert \psi  \rangle _{I N}$ and $\vert \psi  \rangle _{O U T}$, namely on the oracle problem. 

Each partial measurement projects the initial quantum superposition of all the
possible problem-settings, namely $\sum _{b \in \sigma }\vert b \rangle _{B}\vert  0 \ldots  \rangle _{A}$, on the smaller superposition $\sum _{b \in \sigma ^{ \prime }}\vert b \rangle _{B}\vert  0 \ldots  \rangle _{A}$, an instance of Alice's advanced knowledge (knowledge that the problem-setting $b$ belongs to $\sigma ^{ \prime }$).

It is easy to see that this method is equivalent to the one adopted in the previous sections in the case of
Grover algorithm. Here any measurement of the content of register $A$ in the output state is equivalent to the corresponding measurement of the content of register $B$; moreover, measuring any content of $B$ in the input or output state is the same since the reduced density operator of $B$ does not change along the unitary transformation in between. Therefore both partial measurements can be performed in the input
state. 

The same is evidently true in the case that the solution is a one to one function of the problem-setting. If it is not, in
the output state of the unitary part of the quantum algorithm to Alice, a superposition of problem-settings multiplies each solution. However, the final measurement of the
solution anyhow projects this superposition onto the problem-setting selected by Bob. This is because it also triggers the projection due to the initial
measurement postponed at the end of the unitary part of Alice's action. All is as if Alice eventually measured both the problem-setting and the solution.
Anyway we have a one to one correspondence between the initial and final measurement outcomes.

In the following section we apply the
AK rule to the exponential speedups. 

\subsection{ Exponential speedups}
We compare the prediction of the advanced knowledge rule with the major quantum algorithms that yield an exponential speedup: Deutsch\&Jozsa
algorithm (the seminal exponential speedup), Simon algorithm, and Shor factorization algorithm. The results obtained will be discussed in Section 3.3.2.

\subsubsection{ Deutsch\&Jozsa algorithm}
We consider the oracle problem solved by Deutsch\&Jozsa algorithm $\left [29\right ]$. The set of functions known to both Bob and Alice is the set of
the functions $f_{b} :\left \{0 ,1\right \}^{n} \rightarrow \left \{0 ,1\right \}$ that are either constant or \textit{balanced}. In constant functions, the values of the function are of course either all
zeros or all ones. In balanced functions, half of the values are zeros, the other half are ones. For example, for $n =2$, the argument of the function $a$ ranges over $\left \{00 ,01 ,10 ,11\right \}$ and the value of the function $f_{b} \left (a\right )$ over $\left \{0 ,1\right \}$; there are in total eight possible constant and balanced functions,
the following table gives four of them. The first vertical column on the left gives the values of $a$; the corresponding values of each function are in the other columns: 

\begin{equation}\begin{array}{cccccc}a & f_{0 0 0 0} \left (a\right ) & f_{1 1 1 1} \left (a\right ) & f_{0 0 1 1} \left (a\right ) & f_{1 1 0 0} \left (a\right ) & \ldots  \\
00 & 0 & 1 & 0 & 1 & \ldots  \\
01 & 0 & 1 & 0 & 1 & \ldots  \\
10 & 0 & 1 & 1 & 0 & \ldots  \\
11 & 0 & 1 & 1 & 0 & \ldots \end{array}
\end{equation}

Note that we chose as problem-setting $b$ (the bit string suffix of the function) the table of the function, namely from left to right the sequence of function values
for increasing values of the argument. In this way the problem-setting maintains the structure of the problem in it. 

Bob selects a
valuation of $b$ out of all the possible valuations and gives Alice the black box that computes the function $f_{b} \left (a\right )$. Alice is to find whether the function selected by Bob is constant or balanced by performing function evaluations for suitable values
of the argument $a$. 

In the classical case, and in the worst case, the number of function evaluations required to solve the
problem grows exponentially with $n$. In the quantum case just one function evaluation is required. 

The particular structure of the present
problem (reflected in the form of the problem-setting) allows an easy application of the AK rule. 

Let us assume that Bob's measurement
of $\hat{B}$ selects the problem-setting $b =0011$. This measurement can be split into only a pair of partial measurements that satisfy the AK rule. In fact, if one partial
measurement selected both zeros and ones, it would already identify the solution, namely the fact that the function is balanced. Then the cases are two:
If also the other identified the solution, there would be redundancy between the selections performed by the two partial measurements. If it did not, then
the two partial measurements would not evenly contribute to the selection of the solution. In either case there would be a violation of the AK rule. 

Therefore, one partial measurement should select the two left digits of $b =0011$ (both zero), the other the two right digits (both one). In the former case, Alice knows in advance that $f_{0 0 1 1} \left (00\right ) =0$ and $f_{0 0 1 1} \left (01\right ) =0$. Of course she can find whether the function is constant or balanced with just one function evaluation, for any value of the
argument not in the half table she knows in advance. The discussion of the latter case is of course completely similar. 

The above
holds of course for any value of $n$. Alice must know in advance a \textit{good half table}, namely $2^{n -1}$ rows of the table in which all the values of the function are the same. Then she can always solve the problem with just one function
evaluation, for any value of the argument outside the part of the table she knows in advance. 

Summing up, the AK rule says that Deutsch\&Jozsa's
problem can be solved quantumly with just one function evaluation. This is the number required by Deutsch\&Jozsa algorithm, which is of course optimal
since less than of one function evaluation is impossible. 

\subsubsection{ Simon algorithm}
Now we consider the oracle problem solved by Simon algorithm $\left [30\right ]$. The set of functions are the ``periodic'' functions $f_{b} :\left \{0 ,1\right \}^{n} \rightarrow \left \{0 ,1\right \}^{n -1}\text{}\text{}$ such that $f_{b} \left (a\right ) =f_{b} \left (c\right )\text{}$ if and only if $c =a$ or $c =a \oplus p_{b}$ . Here the symbol $ \oplus $ denotes bitwise modulo 2 addition; $p_{b}$, depending on $b$, is a sort of period of the function. For example, see below the tables of three of the six functions for $n =2$ (in the tables of the other three, the values of the function $1\text{}\text{}\text{}$ become $0\text{}$ and vice-versa): 

\begin{equation}\begin{array}{cccc}\text{\thinspace \thinspace } & p_{0 0 1 1} =01 & p_{0 1 0 1} =10 & p_{0 1 1 0} =11 \\
a & f_{0 0 1 1} \left (a\right ) & f_{0 1 0 1} \left (a\right ) & f_{0 1 1 0} \left (a\right ) \\
00 & 0 & 0 & 0 \\
01 & 0 & 1 & 1 \\
10 & 1 & 0 & 1 \\
11 & 1 & 1 & 0\end{array} \label{simon}
\end{equation}

Bob selects a valuation of $b$ and gives Alice the black box that computes $f_{b} \left (a\right )$. Alice is to find the period $p_{b}$ by performing function evaluations for suitable values of the argument $a$. 

The number of function evaluations required by the optimal classical algorithm is $\ensuremath{\operatorname*{Exp}} \left (n\right )$, that required by Simon algorithm is $\ensuremath{\operatorname*{O}} \left (n\right )$. There is thus an \textit{exponential speedup}. The unitary part of this quantum algorithm produces, with just one function
evaluation, a superposition of bit strings orthogonal to the bit string $p_{b}$. The final measurement selects one of them at random. By repeating the entire process until finding $n$ different bit strings, what requires on average an $\ensuremath{\operatorname*{O}} \left (n\right )$ number of repetitions, one can find $p_{b}$ by solving a system of linear equations. 

We apply the AK rule directly to the problem of finding the period of
the function $p_{b}$. Also now the problem-setting, the bit string $b$, corresponds to the table of the function $f_{b}$. The two partial measurements should select respectively two even parts of $b$ -- of the table of the function -- in each of which the value of the function does not repeat; otherwise the partial measurement
in question would identify the period (the solution) -- we have seen that this would violate the AK rule. In other words, each partial measurement should
select a part of the table of the function corresponding to one period\protect\footnote{
By the way, being from $\left \{0 ,1\right \}^{n} \rightarrow \left \{0 ,1\right \}^{n -1}$, the functions in question have just two periods.}. For example, let us assume that $b =0011$. The period of $f_{0 0 1 1} \left (a\right )$ is $p_{0 0 1 1} =01$. One partial measurement should select the first and third bit of $b$ from the left, the other the second and fourth. In the former case Alice knows in advance that $f_{0 0 1 1} \left (00\right ) =0$ and $f_{0 0 1 1} \left (10\right ) =1$. She can find the period of the function with just one function evaluation, for any value of the argument outside the part of
the table she knows in advance. She naturally obtains a value of the function already in the part of the table she knows in advance,$\text{}$ what identifies the period. The discussion of the latter case is completely similar. 

Summing
up, the AK rule tells us that an optimal quantum algorithm solves Simon's problem with just one function evaluation, against the $\ensuremath{\operatorname*{O}} \left (n\right )$ of Simon algorithm. This should imply that Simon algorithm is suboptimal\protect\footnote{
It is anyhow almost optimal. in fact a $\ensuremath{\operatorname*{O}} \left (n\right )$ increase in the number of function evaluations does not affect the exponential character of the speedup. }.

We showed that this is in fact the case for $n =2$, where we were able to produce a quantum algorithm that yields the period with just one function evaluation $\left [9\right ]$. 

By the way, as it is, the AK rule could
not be applied directly to the original Simon's problem -- finding a bit string orthogonal to the period -- since in this case the final measurement does
not occur in an eigenstate of the measured observable. There is not the required unitary transformation between the initial and final measurement outcomes.

\subsubsection{ Shor factorization algorithm}
In the quantum subroutine of Shor factorization algorithm $\left [31\right ]$, the problem is to find the period of the function selected by
Bob from a set of properly periodic functions. This requires a number of function evaluations exponential in problem size with the best known classical
algorithm, polynomial with the quantum subroutine in question. There is thus an exponential speedup. 

As in the case of Simon's problem,
the two partial measurements should select respectively two (consecutive) parts of the table of the function corresponding to two periods of it. Thus, Alice
knows in advance a part of the table of the function corresponding to a period. She can identify the period with just one function evaluation for a value
of the argument immediately outside the part of the table she knows in advance. 

In conclusion the AK rule tell us that Shor's problem can be solved
with just one function evaluation. 

There is a difficulty similar to that of Simon algorithm. To find the period, the quantum subroutine
of Shor algorithm must be repeated a number of times polynomial in the size of the problem. According to the AK rule, also Shor algorithm would be suboptimal.

By the way, all the problems examined in this section are instances of the Abelian hidden subgroup problem $\left [32\right ]$. Since these problems are always about finding some kind of period
of a particular type of periodic function, the AK rule might tell us that they are always solvable with just one function evaluation. Checking this possibility
should be for further work. 

\section{ Discussing the physical character of the explanation}
Let us summarize the key steps of our argument: 

i) The usual representation of quantum algorithms is extended to the process
of setting the problem. Bob, by the initial measurement, selects the problem-setting out of a state of complete indetermination of it. 

ii)
If Alice were the observer of the initial measurement, she would know the problem-setting and thus the solution of the problem before starting her problem
solving action. We must shield her from the outcome of the initial measurement. 

iii) However, under a time-symmetric representation
of the reversible process between the initial and final measurement outcomes, this works only partially. Alice remains shielded from the information (about
the problem-setting and the corresponding solution) coming to her forward in time from the initial measurement, not from the that coming to her backwards
in time from the final measurement. All is as if, before performing her problem-solving action, she already knew half of the information about the problem-setting
and the solution she will read in the future and could use this information to compute the solution with fewer function evaluations. 

We
call this situation a \textit{half causal loop}. Whether it is physical is the subject of the present discussion. With this question in mind,
we go through the critical steps of our argument. 

\subsection{ A critique of the notion of quantum computation }
Let us compare the usual and the extended representation of quantum algorithms. We focus on the four drawer instance of Grover's algorithm in
the assumption that the problem-setting (the number of the drawer with the ball) is $b =01$. 

The usual representation of the quantum algorithm is given in table (\ref{usual}). We have
seen that register $V$ can be disregarded; the comparison is easier if we multiply everywhere the quantum state by the constant ket $\vert 01 \rangle _{B}$; correspondingly we can replace $\hat{U}$ by $\hat{\mathbf{U}}$. With this, the usual representation becomes:
\begin{equation*}\begin{array}{ccc}\text{\thinspace \thinspace } & \text{\thinspace \thinspace } & \text{meas.}\hat{A}\text{} \\
\vert 01 \rangle _{B}\vert  00 \rangle _{A} &  \Rightarrow \hat{\mathbf{U}} \Rightarrow  & \vert 01 \rangle _{B}\vert  01 \rangle _{A}\end{array}
\end{equation*}We compare it with the extended representation of table (\ref{ex}). Simply, in the usual
representation, the process of preparing the problem-setting $\vert 01 \rangle _{B}$ (the preparation of $\vert 00 \rangle _{A}$ is irrelevant here) and the corresponding initial measurement are missing. 

This way of reducing the whole process
to a part thereof, replacing the missing part by its outcome $\vert 01 \rangle _{B}\vert  00 \rangle _{A}$ (the outcome of the problem-setting process), which becomes the input of the represented part, would be normal practice in the classical
case. Here the principle of locality -- that an object is directly influenced only by its immediate spatio-temporal surrounding -- allows to break down
the whole into adjacent spatio-temporal parts each with its suitable surrounding conditions. In the classical framework, the consequent notion of algorithm,
or Turing machine, limited to the computational part of problem-solving, remains fundamental in character (a complete description). 

However,
in the quantum framework, this kind of reduction, for the very reason of relying on the principle of locality, can preclude seeing nonlocal quantum effects.
It obviously precludes seeing any temporal nonlocality given that the surviving process involves just the final measurement. We believe that this is a shortcoming
of the very notion of quantum computation. It should be evident in hindsight, but it has never been noticed before. 

\subsection{ Step (i): relativizing the quantum algorithm to the problem solver }
The first step of the present interpretation of the speedup, extending the usual representation of quantum algorithms to the process of setting
the problem, is of course legitimate. However, it inevitably brings us into the little charted waters of relational quantum mechanics. 

It is convenient to start our discussion
from the abstract process that should be physically represented. It is a process of game theory. The latter can be defined as \textit{the study of the
mathematical models of conflict and cooperation between rational decision makers} (or \textit{agents}). In the case of the four drawers
problem, the decision makers are Bob, who selects the value of $b$ (the number of the drawer with the ball) and Alice, who is to identify that value by opening the minimum number of drawers.
It is essential that the value of $b$ selected by Bob is hidden from Alice. Otherwise she would know the solution of the problem without opening any drawer. In technical
jargon, to Alice, the value of $b$ must be hidden inside a \textit{black box}. 

Quantum computation is the physical representation
of the abstract notion of computation, here the computation of the solution of the problem. Moving from quantum computation to quantum problem-solving implies
physically representing the above game situation. 

We should find in the first place the physical correspondent of the decision makers.
It is natural to assume that they become the quantum mechanical observers.

Bob,
the decision maker who selects the number of the drawer with the ball, naturally becomes the observer of the initial measurement, who is in control of register $B$ and does the same by measuring its content. Alice, the decision maker who locates the ball by opening drawers, becomes the observer
of the final measurement, who is in control of register $A$ and does the same by performing function evaluations and eventually measuring the content of $A$. 

When we say that $\vert 01 \rangle _{B}\vert  00 \rangle _{A}$ cannot be the input state to Alice because it would tell her the solution of the problem, we simply mean that Alice cannot be the observer
of the initial measurement. To her, the state after the initial Bob's measurement must remain one of complete ignorance of the problem-setting selected by Bob, namely the unprojected state $(\vert 00 \rangle _{B} +\vert 01 \rangle _{B} +\vert 10 \rangle _{B} +\vert 11 \rangle _{B})\vert  00 \rangle _{A}$. This is physically represented by postponing the projection
of the quantum state due to the initial Bob's measurement to the end of the unitary part of her problem-solving action or, indifferently, by postponing
the initial measurement itself.

By the way, the above might have an implication on the notion of observer. Whether one can talk about the observer as if it were just the pointer of the measurement apparatus or an observer we can identify ourselves with is a well known subject of debate. It would seem that the need of relativizing the quantum algorithm to Alice implies the latter option. However, we would not go into this further. We simply note that a similar problem applies to the game theory notion of ``rational agent'': can it be a machine or should it be an anthropomorphic agent? Our point would be that, if we want to move forward in the physical representation of abstract information notions, which should be the purpose of quantum information, we must accept to incorporate this problem.

The above relativization of the quantum state to Alice naturally implies giving up the notion of absolute quantum state of the Copenhagen interpretation and entering into a relational form of quantum mechanics. Among these we find the \textit{relational interpretation} $[33 ,34]$, \textit{quantum Bayesianism} (QBism) $[35 ,36]$, and Healey $\left [37\right ]$ pragmatist approach to quantum mechanics. For example $\left [34\right ]$, at a certain time, to one observer the state of the quantum system
may be collapsed on an eigenstate of the measured observable, while to another observer it might still be a superposition of all the eigenstates (this is
exactly our case). 

Of course, the perspectives of all the observers must be coherent with one another according to some criteria $\left [34\right ]$. Presently, this is simply the postponement, with respect to the
observer Alice, of the projection of the quantum state associated with the initial measurement. According to it: 

(I) The initial state
of complete indetermination of the problem-setting, the unitary operator $\hat{\mathbf{U}}$, eventually the overall final state (after the postponed projection) are the same to all observers. 

(II) The
representation of the unitary evolution of the quantum state between the initial and final measurement outcomes depends on the observer: we must have a
representation to the problem setter and any external observer and a different one to the problem solver. 

This means introducing a
notion of perspective in the quantum mechanical description; the description depends on the point of view of the observer. An analogy with geometry can
be useful. The same object appears differently depending on the point of space from which one looks at it, but in fact all these perspectives must be coherent
with one another according to some criteria. 

Summing up, all the above seems to be quite natural when one moves from computation to problem-solving.
Computation is the action of a single agent, and can thus be physically represented within the Copenhagen interpretation and the notion of absolute quantum
state (the same for all observers). Problem-solving is a game theory notion and essentially involves two agents -- the problem setter and the problem solver -- and the notion of hiding from one the information available to the other. Going to its physical representation, it becomes necessary to enter into a relational form of quantum mechanics.

Let us eventually counter a possible objection, that Alice should be shielded not only from the information
coming to her from the initial measurement of $\hat{B}_{l}$, but also from that coming to her from the final measurement of $\hat{A_{r}}$. Interestingly, adopting a shielding symmetric in time with respect to that from the initial measurement would not work. The time-symmetric
thing would be advancing the projection of the quantum state associated with the final measurement of $\hat{A_{r}}$ to the beginning of the unitary part of the quantum algorithm $\hat{\mathbf{U}}$. However, this would not shield Alice from anything, on the contrary it is what creates her advanced knowledge. The only way would
be to omit the final measurement of $\hat{A_{r}}$, but this would disrupt the quantum algorithm. 

\subsection{Step (ii): time-symmetrizing the relativized quantum algorithm}
The time-symmetrization of the quantum algorithm to Alice creates what we have called the \textit{half causal loop}. In retrospect, all is as if Alice
knew in advance half of the information about the problem-setting and the solution she will read in the future and could use this information to reach the
solution with fewer function evaluations. In the following we discuss whether this loop can be physical. 

\subsubsection{ Comparison with the major quantum algorithms}
The half causal loop in question is implicit in the advanced knowledge (AK) rule. It is therefore ``physical'' (in the sense that there is
nothing against this possibility) in the optimal quantum algorithms that require the number of function evaluations predicted by it. 

This
is the case of Grover algorithm, which is demonstrably optimal and requires $\ensuremath{\operatorname*{O}} \left (2^{n/2}\right )$ function evaluations -- where $n$ is problem size. This is also the number predicted by the AK rule. The half causal loop is physical here. This should also be
the most interesting case, since Grover's problem, the unstructured problem par excellence is, so to speak, fundamental in character. 

The
quantum algorithms that yield exponential speedups solve instead highly structured problems. We have examined: Deutsch\&Jozsa algorithm (the seminal
exponential speedup), Simon algorithm, and the quantum subroutine of Shor's factorization algorithm. The AK rule predicts that the corresponding problems
can be solved quantumly with just one function evaluation. This prediction exactly fits Deutsch\&Jozsa algorithm, not Simon's algorithm and the quantum
subroutine of Shor's algorithm, which require respectively $\ensuremath{\operatorname*{O}} \left (n\right )$ and $\ensuremath{\operatorname*{Poly}} \left (n\right )$ function evaluations (with $n$ always problem size). Thus, the latter algorithms would be suboptimal according to the AK rule. We showed that it is actually
so in the trivial case of Simon algorithm with $n =2$ -- here we could produce a quantum algorithm that yields the solution with just one function evaluation. Checking the situation
more in general is likely difficult. Of course, the fact remains that the AK rule is in good agreement also with these quantum algorithms for values of
$n$ such that $\ensuremath{\operatorname*{Poly}} \left (n\right )/\ensuremath{\operatorname*{Exp}} \left (n\right ) \approx 0$. 

Our conclusion would be that the AK rule accords well with all the above quantum algorithms, which are
the major ones. Furthermore, it provides a good insight into the reason for their speedup. 

The advanced knowledge of half of the problem-setting
clearly explains the quadratic speedup when the problem is to identify the number of the drawer with the ball. In fact, the advanced knowledge of $n/2$ of the bits that specify it reduces the number of drawers to search the ball in from $2^{n}$ to $2^{n/2}$. 

When the problem must be solved classically with an $\ensuremath{\operatorname*{Exp}} \left (n\right )$ number of function evaluations and is structured in such a way that knowing half of its setting in advance allows to find the solution
with just one function evaluation, of course the AK rule explains the exponential speedup of the quantum case. 

In hindsight, one would
say that the problems liable of being solved with a quantum speedup have been selected with the AK rule in mind. 

\subsubsection{ Discussing the form of time-symmetrization adopted}
We discuss the form of time-symmetrization of the quantum algorithm adopted. 

We put ourselves in the simplest case that
the solution is a one to one function of the problem-setting: so the information that specifies one specifies also the other. We should keep in mind that
there is a unitary transformation between the initial measurement outcome, which selects the problem-setting in a state of complete indetermination of it,
and the final measurement outcome, which selects the corresponding solution. Since no information is destroyed along it, the process between the two outcomes
is physically reversible. We thus require that its representation is symmetric with respect to time. 

This requirement has been satisfied
by assuming that the selection of the information that specifies the problem-setting (or identically the solution) evenly and non-redundantly shares between
the initial and final measurements in a quantum superposition of all the possible ways of doing this. In the following, we discuss the assumption in question
from a variety of points of view, not excluding the psychological one. 

\textbf{\strut } \smallskip 

\textbf{\emph{Perspective 1}} 

The assumption that the selection of the information that specifies
either the initial or final measurement outcome evenly shares between the initial and final measurements can also be justified in a way that is not related
to time. 

We have seen that the initial measurement of $\hat{B}$ (the content of register $B$) can be postponed at will along the unitary transformation $\hat{\mathbf{U}}$ which represents Alice's problem-solving action. In fact, this action never changes the problem-setting selected by Bob; thus the reduced
density operator of register $B$ remains unaltered along it. Moreover, $\hat{B}$ and $\hat{A}$ commute. Therefore, we can postpone the measurement of $\hat{B}$ to the time of the final measurement of $\hat{A}$ (the content of register $A$). In this way the two measurements are simultaneously performed in the state of maximum entanglement in the top-right corner
of table (4). Note the perfect symmetry between these two measurements in the state in question: nothing changes
if we interchange the labels $B$ and $A$ of both the observables and the quantum states. 

Now there is no more the justification of ascribing the
selection of the information that specifies either the initial or the final measurement outcome to the initial measurement because it is performed first.
However, anyhow Occam's razor always requires that we get rid of the redundancy between the two selections. The only way of doing this while
satisfying the symmetry in question is to reduce the two measurements to partial measurements that evenly contribute to the selection in a uniform superposition
of all the possible ways of evenly sharing the information. This is indeed the form of time-symmetrization adopted. 

In a way, causality that goes backwards
in time arises when the two simultaneous measurements are moved away from each other along the time dimension. For a comparison between the backward causality
picture and the simultaneous one in quantum mechanics in general, see Adlam $\left [38\right ]$. 

\medskip \textbf{\emph{Perspective
2}} 

Let us compare the assumption of evenly sharing the selection of the information between the initial and final
measurements with the other alternatives. Alternative \#1 is that the information is all selected by the initial measurement. Alternative \#2 that it is all
selected by the final measurement. We propose an argument that would exclude these extremal alternatives. 

Under \#2, the zigzag diagram
of the quantum algorithm to Alice of table (\ref{sya}) would tell her in advance the whole setting/solution of the problem.
This would clearly be unphysical: a complete causal loop. However, since alternative \#1 is the time-symmetric image of \#2 with the same right to existence given that the process is reversible, it should be excluded as well. So to speak, while in \#2 Alice would know in advance too much, in \#1 she would know too
little, indeed nothing at all about the problem-setting even if there can be a speedup. 

Of course, the only way to avoid the two extremal
alternatives is to properly share the selection of the information between the initial and final measurements. 

\medskip \textbf{\emph{Perspective 3}} 

Let us look at the zigzag diagram of table (\ref{sya}).
The bottom line of the diagram, which inherits both selections, is the time-symmetrized quantum algorithm (one of the instances thereof). In the left end
of it, the state of register $B$ is $\vert 01 \rangle _{B} +\vert  11 \rangle _{B}$. This means that Alice knows in advance, before beginning her problem-solving action  $\hat{\mathbf{U}}$, that the right bit of $b$ is $1$. Apparently, this information, hosted in register $B$ at the beginning of  $\hat{\mathbf{U}}$, came backwards in time from the final measurement of $\hat{A_{r}}$ -- see table (\ref{sya}). 

Of course, if an information coming from the future
could be measured, Alice's advanced knowledge would be unphysical. The point is, it cannot without breaking the half causal loop. To measure it, Alice should
measure $\hat{B}_{r}$ (the right bit of register $B$) immediately after the initial measurement. Indeed this would tell her that the bit in question is $1$ (we are under the assumption that the initial Bob's measurement selected $b =01$). However, whether the information in question comes backwards in time from the final measurement or forward in time from the
measurement of $\hat{B}_{r}$ could not be distinguished any more. The half causal loop would be broken. 

\medskip \textbf{\emph{Perspective
4}} 

When dealing with retrocausality, the reason one does not see violations of causality, even though apparently
there is information going backwards in time from the final to the initial measurement, has been often attributed to quantum indeterminacy $[39 ,40]$, which also accounts for spatial nonlocality $\left [41\right ]$. In the present case, each time-symmetrization instance is originated
by the backwards in time propagation of a selection performed by the final measurement. However, this comes with its own remedy: having to take the superposition
of all the possible instances. Indeed this superposition -- the unsymmetrized quantum algorithm of table (4)
back again -- is the usual unitary transformation of a quantum superposition, a completely ordinary thing in quantum mechanics. Summing up, in the present
case, the quantum indeterminacy that allows an apparent flow of information backwards in time would be that inherent in the very notion of quantum superposition.

\medskip \textbf{\emph{Conclusion}} 

We are of course in uncharted
waters, but it would seem that several features nicely go together: (i) The time-symmetrization of the quantum algorithm adopted can be justified also
outside time where it becomes the only way of satisfying a clear geometrical symmetry. (ii) In the relational context, it would be unphysical not to properly
share the selection of the information between the initial and final measurements. (iii) There is no observable information sent back in time. (iv) Backward
causality is compensated for by the indeterminacy inherent in the very notion of quantum superposition. 

\subsubsection{ Quantum entanglement}
Looking for a relation between quantum speedup and entanglement is a main direction of research in literature. Its motivation can be traced back
to an observation of Jozsa $\left [4\right ]$. If the state of the quantum computer register always remains the
product of the states of its individual qubits (i.e. unentangled), the quantum algorithm can be efficiently -- in polynomial time -- simulated in a classical
way. Therefore a quantum algorithm without entanglement generation could not give an exponential speedup.

However, until now, no unifying relation between the various exponential
speedups and entanglement has been found. An authoritative opinion about the situation has been given in $\left [6\right ]$. After examining the role played by entanglement in a variety of
quantum algorithms, the author comes to a conclusion that highlights the open character of this research: \textit{we should give up looking for a single
reason behind the quantum speedup. Most likely, the answer will intimately be connected with the exact nature of the problem and, as seen above, will vary
from problem to problem. Though possibly intellectually displeasing, this answer is the only possible consistent one at present. } 

Entanglement plays an essential role also in the present interpretation of the speedup, but in a different way. The entanglement studied in literature
is internal to register $A$ -- is between the contents of the individual quantum bits of this register. The one we are dealing with in the present work
is between the contents of registers $B$ and $A$ (see the top-right corner of table \ref{sya}).

By the way, let
us note that the coexistence of the two forms of entanglement does not endanger entanglement monogamy. In fact, the entanglement between the contents of
registers $B$ and $A$ is at another hierarchical level. It concerns a superposition of quantum algorithms, each for one of the possible problem-settings.
In each of them, while the entanglement internal to register $A$ builds up, the state of register $B$ (which contains the problem-setting) is always the same and factorizable with respect to that of $A$.

 Another difference is that the entanglement between the contents of
registers $B$ and $A$ becomes maximum
simply by reaching the solution of the problem, no matter whether it is reached with a speedup. Alice's advanced knowledge, which depends on the entanglement
in question, only says that the solution can be reached with a speedup. It takes an optimal quantum algorithm to benefit from it. 

For
the time being, the relation between the entanglement appearing in this work and the one addressed in the literature remains unclear. 

Let us note instead an analogy between the current form of temporal nonlocality, which depends on the entanglement between registers $B$ and $A$, and spatial nonlocality.

First, we make a specification. The term \textit{spatial nonlocality}
is used in two meanings: (i) in relation to the fact that a measurement \textit{here }nonlocally (instantly) changes the state the quantum system
\textit{there} (thus also the result of a subsequent measurement on it) and (ii) in relation to the violation of the Bell inequalities on the part of a pair of measurements (a third
meaning which is seemingly irrelevant here is the Aharonov-Bohm nonlocality). 

The temporal nonlocality inherent in the present interpretation
of the speedup is nonlocal action in time. It occurs in each element of the superposition of the time-symmetrization instances. For example, the final measurement of $\hat{A_{r}}$ nonlocally changes the previous (input) state of register $B$ $\left [39\right ]$ -- see the bottom line of table (\ref{sya}).
The point we would like to make is that this form of temporal nonlocality appears to be tightly related to spatial non-locality. 

If
we assumed that registers $B$ and $A$ had been spatially separated just before the final measurement of $\hat{A_{r}}$, then this measurement, besides that of register $A$, would instantly change also the state of the space-separated register $B$. Costa de Beauregard's $\left [12\right ]$ zigzag explanation of this nonlocal action in space would overlap
with the right part of the zigzag diagram of table (\ref{sya}), which highlights instead the temporally nonlocal character
of the speedup. Looking for possible unifications between spatial and temporal nonlocality on the basis of their Parisian zigzag's seems to be an interesting
research prospect. 

\subsubsection{ Temporal Bell inequalities}
Morikoshi $\left [42\right ]$ showed that Grover's algorithm violates an information theoretic
temporal Bell inequality that he specifically developed for quantum computation. By the way, he also noted that this might hint at a temporally nonlocal
character of the speedup. Now, there must be some relation between Alice's advanced knowledge and the violation in question. In fact, Grover algorithm, by
exploiting the former, meets that violation. Investigating the relation between the temporally nonlocal character of the speedup and the violation of temporal
Bell inequalities might be an interesting research prospect. 

\subsubsection{ Retrocausality}
Extending the representation of quantum algorithms to the process of setting the problem naturally brings retrocausality into play. In fact, the
problem-setting and problem solution, in their quantum version, constitute pre- and post-selection, respectively. Hence the process as a whole is bound
to be affected by both boundary conditions. We shall now discuss in more detail the notions of time and causality employed in the present interpretation
of the speedup. 

To start with, let us position in literature the framework that we have employed. It is closely related to the retrocausal
interpretations of quantum mechanics $\left [13 -25\right ]$. In particular, we would mention the Two-State-Vector Formalism
$\left [18 ,25\right ]$. According to this formulation, at any intermediate moment $t$ the full specification of a quantum system is given by the two state $ \langle \phi  (t)||\psi  (t) \rangle $, where $\vert \psi  (t_{i}) \rangle $ is the forward evolving initial state of the system (preselected at time
$t_{i}$), and $ \langle \phi  (t_{f})\vert $ is the backward evolving final state of the system (postselected at time $t_{f}$). The latter can be also thought of as a ``destiny'' state $\left [23\right ]$, propagating towards the future like the initial state and complementing
the information it carries (this is the main ingredient we need for our analysis in this work). It was shown that quantum uncertainty masks any causality
violations which could have arisen in such approach $\left [40 ,41\right ]$ (and hence our analysis of the speedup did not encounter any causality
paradoxes). This formalism of quantum mechanics is known to be equivalent to its standard formulations $\left [18\right ]$, yet over the years it has inspired the discovery of many new quantum
phenomena $\left [20 -25\right ]$. 

For what concerns the notion of time, we
work here within non-relativistic quantum mechanics, where time can be most simply regarded as a parameter. The assumption that causality can move both
forward and backwards in time, might make it intuitive to regard time also as medium through which causal effects can propagate. It should be noted however that these forward and backwards in time propagations should occur according the strict rules of time-symmetric quantum mechanics.

\section{ Conclusion}
Let us go back for a moment to the origin of the quantum computation notion. 

In 1968, Finkelstein $\left [43 ,44\right ]$ demonstrated the feasibility of computation in the quantum world
and introduced the notion of quantum bit. This was not to devise a quantum computer but to represent spacetime as a quantum computer. 

In
1982, Feynman $\left [45\right ]$ addressed the problem of simulating a quantum evolution by a universal
(classical) computer. Interestingly, he explicitly distinguished between two cases: (1) the state of the quantum system depends both on the future and the
past. (2) the state depends only on the past, like in a standard causal evolution. It may be no coincidence that he was, with Wheeler, the author of the
electromagnetic absorber theory. He noted that, in the former case, one would need a cellular automaton subjected to two boundary conditions simulating
the initial and final boundary conditions of the quantum evolution, in the latter an ordinary computer performing an initial-value type
of computation. Then he focused on the latter case showing that the classical simulation of a quantum evolution governed by the Schr{\"o}dinger equation
would have generally required an exponentially higher amount of physical resources. From this he inferred that a quantum computer could be more efficient
than a classical computer in performing the same computational task. Of course Feynman's 1982 paper also availed itself of the notion of (classical) reversible
computation developed by Bennett $\left [46\right ]$, and Fredkin\&Toffoli $\left [47\right ]$. 

Later, in 1985, Deutsch $\left [48\right ]$ devised the seminal quantum algorithm. The quantum computation
of the solution of an elementary oracle problem, a simple unitary evolution, required just one oracle query (function evaluation) against a minimum of two
with a universal (i.e. classical) computer. As is well known, this started modern quantum computation. The subsequent research in the field led to the
discovery of dozens of quantum algorithms, most notably the integer factorization algorithm of Shor (1994) and the search algorithm of Grover (1996). 

The present approach to the quantum
speedup can be made to start from a critique of the notion of quantum computation. Logically, the computation of the solution of the problem must be preceded by the setting of the problem. But this part is omitted in the usual representation
of quantum algorithms, which starts from the input state of the unitary evolution that solves the problem. This would not be a shortcoming in the classic
case. Here the principle of locality allows to break down the whole into adjacent spatio-temporal parts each with its suitable surrounding conditions (also
each part can have a complete description). However, in the quantum framework, this kind of reduction can originate an incomplete description. For the very
reason of relying on the principle of locality, it can preclude seeing nonlocal quantum effects. We argue that this is a shortcoming of the very notion
of quantum computation. It should be evident in hindsight, but as far as we know it has never been noticed before. 

Completing the
representation of quantum algorithms with the process of setting the problem introduces a new perspective under which to see them. 

In the first
place, it highlights the relevance of time-symmetric quantum mechanics to quantum computation: the problem-setting and problem solution, in their quantum
version, constitute pre- and post-selection, hence the process as a whole is bound to be affected by both boundary conditions. Since in the present case
the process between the initial and final measurement outcomes is reversible, for reasons of time-symmetry we should assume that it is affected in an even
way by the initial and final measurements. The selection of the problem-setting and the consequent solution should evenly
and non-redundantly share between them. The selection ascribed to the initial measurement should propagate forward in time, that ascribed to the final measurement backwards in time, this in a quantum superposition of all the possible ways of evenly sharing. With this, one time-symmetrizes the quantum algorithm.

In the second place, it forces us to enter into relational quantum mechanics. There must be a representation of the quantum algorithm
with respect to Bob (the problem setter) and any external observer, and another one with respect to Alice (the problem solver) from whom the problem-setting
selected by Bob with the initial measurement (thus the corresponding solution) should be concealed. 

The synthesis of the above two points is as follows. Time-symmetrizing the quantum algorithm to take into
account both boundary conditions is without consequences in the case of the representation to Bob and any external observer, who are not shielded from any measurement outcome. Instead it shows that the representation to Alice, who is shielded from the outcome of the initial measurement, is a superposition
of time symmetrization instances in each of which she remains shielded from the half information coming to her from the initial measurement, not from the
half coming to her backwards in time from the final measurement. In retrospect, the process of setting and solving the problem looks as if the problem solver
knew in advance, before beginning her problem-solving action, half of the information that specifies the setting and thus the solution of the problem she
will read in the future and could use this information to compute the solution with fewer function evaluations. 

This interpretation
of the speedup provides the number of function evaluations required to solve an oracle problem in an optimal quantum way. It is the number required by a
universal computer (e.g. a Turing machine) that benefits from the advanced knowledge of half of the information about the setting and the corresponding
solution of the problem. This number fits all the quantum algorithms examined, which comprise the major ones and cover both the quadratic and exponential
speedups. 

The affinity between problem-solving and time-symmetric quantum mechanics may have some non-trivial consequences. Consider
the ``weak values'' $\left [49\right ]$ derived by TSVF. These are values that prevail between the pre-
and post-selections, sometimes possessing unique properties. Suppose, e.g., that we measure two non-commuting variables. The weak values prevailing during
the time-interval between these two measurements maintain both the past and future measurement outcomes, together. This makes the information about the
intermediate state greater than what is considered to be allowed by the uncertainty principle, and yet the principle is not violated, because this double
knowledge is retrospective. Moreover, under special combinations of pre- and post-selections, where the two outcomes are unlikely to pair, the weak values
become odd $\left [20\right ]$, e.g., too large/small or even complex. Some surprising predictions
based on this fleeting existence of odd values, yielding, for instance, disappearance and reappearance of particles across distant locations $\left [50\right ]$, can be inferred even with the aid of ordinary, projective quantum
measurements $[50 ,51]$ . The bearing on the above analysis of quantum algorithms is again straightforward:
Odd physical values must be part and parcel of many quantum computations. 

All this calls attention to the fact that quantum computation,
while rigorously predicted to surpass classical computation in many cases, is still lagging with respect to some fundamental questions, e.g. what makes
it so effective? While quantum computation proceeds along a well-defined algorithm, the method for finding such algorithms is still a matter of mathematical
ingenuity or even genius, still eluding general formalization. One can therefore pose the challenge for quantum computation to search for such a ``super-algorithm'',
namely asking, ``what kind of algorithm is needed for making a certain computation?'' Such a search, perhaps, can gain new insights from extending the
physical representation from computation to problem-solving and the TSVF. Indeed one important feature of such complex processes is that they often involve
brief ``local inconsistencies'', which have to be tolerated until the final solution proves contradiction-free. This tolerance of ambiguity and apparent
inconsistencies may be the key for quantum computation's advantage over the classical one. 

The current paper hopefully lays the foundation
for further work. We shall conclude with a few future prospects. 

On the quantum computing front, one could extend the verification
of the advanced knowledge rule to other quantum algorithms (there are many by now). Another interesting venue would be to apply this rule to assess the
computational complexity of oracle problems, identify classes of problems and compare them with the known ones. 

As for the foundations
of quantum mechanics: We should have highlighted the importance of physically representing information processes that are logically complete -- otherwise
the quantum description itself might be incomplete. 

In addition, we showed that the physical representation of new abstract information notions might
be an interesting direction of research. We can already position along it the physical representation of the notion of communication, which led to the discovery of quantum cryptography by Bennett\&Brassard
$\left [52\right ]$ and Ekert $\left [53\right ]$, and that of the notion of computation, which led to the discovery of the quantum computational speedup by Deutsch. However, the fact that quantum information is also called quantum communication and computation,
which would confine it to the physical representation of only these two notions, apparently shows that we are dealing with a research direction that went unnoticed so
far. 

Going from the physical representation of computation to that of problem-solving, in particular of the game theory notion of hiding from a player the information available to another player, would
be another step in this direction. It leads to the interplay between the relational and time-symmetric interpretations of quantum mechanics that
seems to elucidate the quantum speedup. 

Historically, the advancement of physics was driven bottom-up by the necessity of representing
new experimental facts. The physical representation of new abstract information notions might be a complementary, top-down, driver (see also $\left [24\right ]$). Besides highlighting new physical effects, it might shed light on the foundations of quantum mechanics. 

Of course, this also means entering uncharted waters. An
important point is that this was unavoidable in the present case. Extending the representation of quantum algorithms to the problem-setting process is necessary in order not to prevent the visibility of quantum nonlocality. This, in turn, forces us to enter into the interplay between the relativization
and time-symmetrization of quantum processes. 

\section*{Acknowledgements}
We wish to thank Yakir Aharonov and David Ritz Finkelstein for many helpful discussions. 

\section{References}
$\left [1\right ]$ Grover, L. K.: A fast quantum mechanical algorithm for database
search. Proc. 28th Annual ACM Symposium on the Theory of Computing. ACM press New York 212-219 (1996)  \\\relax $\left [2\right ]$ Mosca, M. and Ekert, A. K.: The Hidden Subgroup Problem and Eigenvalue
Estimation on a Quantum Computer. Lecture Notes in Computer Science, Vol.1509 (1999)  \\\relax $\left [3\right ]$ Ambainis, A.: Understanding Quantum Algorithms via Query Complexity.
arXiv: 1712.06349 (2017)  \\\relax $\left [4\right ]$ Jozsa, R.: Entanglement and Quantum Computation. Geometric Issues
in the Foundations of Science, Oxford University Press (1997) arXiv:quant-ph/9707034  \\\relax $\left [5\right ]$ Ekert, A. K. and Jozsa, R.: Quantum Algorithms: Entanglement Enhanced
Information Processing arXiv:quant-ph/9803072 (1998)  \\\relax $\left [6\right ]$ Vedral, V.: The elusive source of quantum effectiveness. Found.
Phys., Vol 40, Issue 8, 1141-1154 (2010)  \\\relax $\left [7\right ]$ Aaronson, S. and Ambainis, A.: Forrelation: a Problem that Optimally
Separates Quantum from Classical Computing. arXiv:1411.5729 (2014)  \\\relax $\left [8\right ]$ Castagnoli, G. and Finkelstein, D. R.: Theory of the quantum speedup.
Proc. Roy. Soc. A 1799, 457, 1799-1807 (2001)  \\\relax $\left [9\right ]$ Castagnoli, G.: The quantum correlation between the selection of
the problem and that of the solution sheds light on the mechanism of the quantum speed up. Phys. Rev. A 82, 052334 (2010)  \\\relax $\left [10\right ]$ Castagnoli, G.: Completing the Physical Representation of Quantum
Algorithms Provides a Quantitative Explanation of Their Computational Speedup. Found. Phys. 48, 333-354 (2018)  \\\relax $\left [11\right ]$ Von Neumann, J.: Mathematical Foundations of Quantum Mechanics.
Princeton University Press (1955, 2018)  \\\relax $\left [12\right ]$ Costa De Beauregard, O.: The 1927 Einstein and 1935 EPR paradox.
Physics A 2, 211-242 (1980)  \\\relax $\left [13\right ]$ Dolev, S. and Elitzur, A. C.: Non-sequential behavior of the wave
function. arXiv:quant-ph/0102109 v1 (2001)  \\\relax $\left [14\right ]$ Elitzur, A.C., Cohen, E.: Quantum oblivion: A master key for many
quantum riddles. Int.J. Quant. Inf. 12, 1560024 (2015)  \\\relax $\left [15\right ]$ Elitzur, A. C. and Cohen, E.: 1-1 = Counterfactual: on the potency
and significance of quantum non-events. Phil. Trans. R. Soc. A 374, 20150242 (2016)  \\\relax $\left [16\right ]$ Wheeler, J. A. and Feynman, R. P.: Interaction with the Absorber
as the Mechanism of Radiation. Rev. Mod. Phys. 17, 157-181 (1945)  \\\relax \noindent $\left [17\right ]$ Watanabe, S.: Symmetry of physical laws. Part III. Prediction and
retrodiction. Rev. Mod. Phys. 27 (2), 179-186 (1955)  \\\relax \noindent $\left [18\right ]$ Aharonov, Y., Bergman, P. G., and Lebowitz, J. L.: Time Symmetry
in the Quantum Process of Measurement. Phys. Rev. 134, 1410-1416 (1964)  \\\relax $\left [19\right ]$ Cramer, J.: The Transactional Interpretation of Quantum Mechanics.
Rev. Mod. Phys. 58, 647 (1986)  \\\relax $\left [20\right ]$ Aharonov, Y. and Rohrlich, D.: Quantum paradoxes. Wiley-VCH, Weinheim
(2005)  \\\relax $\left [21\right ]$ Aharonov, Y. and Vaidman, L.: The Two-State Vector Formalism: An
Updated Review. Lect. Notes Phys. 734, 399-447 (2008)  \\\relax $\left [22\right ]$ Aharonov, Y., Colombo, F., Popescu, S., Sabadini, I., Struppa,
D. C., and Tollaksen, J.: Quantum violation of the pigeonhole principle and the nature of quantum correlations. Proc. Natl. Acad. Sci. 113, 532-535 (2016)
\\\relax $\left [23\right ]$ Aharonov, Y., Cohen, E., and Landsberger, T.: The two-time interpretation
and macroscopic time-reversibility, Entropy 19, 111 (2017)  \\\relax $\left [24\right ]$ Aharonov, Y., Cohen E., and Tollaksen, J.: Completely top-down
hierarchical structure in quantum mechanics. Proc. Natl. Acad. Sci. USA 115, 11730-11735 (2018)  \\\relax $\left [25\right ]$ Aharonov, Y, Cohen, E., Carmi, A., and Elitzur, A. C.: Extraordinary
interactions between light and matter determined by anomalous weak values. Proc. Roy. Soc. A 474, 20180030 (2018)  \\\relax $\left [26\right ]$ Bennett, C. H., Bernstein, E., Brassard, G., and Vazirani, U.:
Strengths and Weaknesses of Quantum Computing. SIAM Journal on Computing 26, 1510-1523 (1997)  \\\relax $\left [27\right ]$ Long, G. L.: Grover algorithm with zero theoretical failure rate.
Phys. Rev. A 64, 022307-022314 (2001)  \\\relax $\left [28\right ]$ Toyama, F. M., van Dijk, W., and Nogami Y.: Quantum search with
certainty based on modified Grover algorithms: optimum choice of parameters. Quant. Inf. Proc. 12, 1897-1914 (2013)  \\\relax $\left [29\right ]$ Deutsch, D. and Jozsa, R.: Rapid solution of problems by quantum
computation. Proc. Roy. Soc. A 439, 553-558 (1992)  \\\relax $\left [30\right ]$ Simon, D.: On the power of quantum computation. Proceedings of
the 35th Annual IEEE Symposium on the Foundations of Computer Science 116-123 (1994)  \\\relax $\left [31\right ]$ Shor, P.: Algorithms for quantum computation: Discrete log and
factoring. Proceedings of the 35th Annual IEEE Symposium on the Foundations of Computer Science 124-131 (1994)  \\\relax $\left [32\right ]$ Kaye, P., Laflamme, R., and Mosca, M.: An Introduction To Quantum
Computing. Oxford University Press 146-147 (2007)  \\\relax $\left [33\right ]$ Rovelli, C.: Relational Quantum Mechanics. Int. J. Theor. Phys.
35, 637-658 (1996)  \\\relax $\left [34\right ]$ Rovelli, C.: Relational Quantum Mechanics (2011) http://xxx.lanl.gov/pdf/quant-ph/9609002v2
\\\relax $\left [35\right ]$ Fuchs, C. A.: On Participatory Realism. arXiv:1601.04360v3 [quant-ph]
(2016)  \\\relax $\left [36\right ]$ Fuchs, C. A.: QBism, the Perimeter of Quantum Bayesianism. arXiv:1003.5209v1
[quant-ph] (2010)  \\\relax $\left [37\right ]$ Healey, R.: Quantum Theory: a Pragmatist Approach. arXiv:1008.3896
(2010)  \\\relax $\left [38\right ]$ Adlam, E.: Spooky Action at a Temporal Distance. Entropy 20 (1):
41 (2018)  \\\relax $\left [39\right ]$ Aharonov, Y., Cohen, E., and Elitzur, A. C.: Can a future choice
affect a past measurement outcome? Ann. Phys. 355, 258-268 (2015)  \\\relax $\left [40\right ]$ Aharonov, Y., Cohen, E., and Shushi, T.: Accommodating Retrocausality
with Free Will. Quanta 5, 53-60 (2016)  \\\relax $\left [41\right ]$ Carmi, A. and Cohen, E.: Relativistic independence bounds nonlocality.
Sci. Adv. 5, eaav8370 (2019)  \\\relax $\left [42\right ]$ Morikoshi, F.: Information-theoretic temporal Bell inequality and
quantum computation. Phys. Rev. A 73, 052308 (2006)  \\\relax $\left [43\right ]$ Finkelstein, D. R.: Space-time structure in high energy interactions.
Gudehus, T., Kaiser, G., Perlmutter, A. editors. Conference on high energy interactions, Coral Gables (1968)  \\\relax $\left [44\right ]$ Finkelstein, D. R.: Space-Time Code. Phys. Rev. 184, 1261 (1969)
\\\relax $\left [45\right ]$ Feynman, R. P.: Simulating Physics with Computers. Int. J. Theor.
Phys. 21, 467-488 (1982)  \\\relax $\left [46\right ]$ Bennett, C. H.: The Thermodynamics of Computation -- A Review.
Int. J. Theor. Phys. 21, 905-940 (1982)  \\\relax $\left [47\right ]$ Fredkin, E. and Toffoli, T.: Conservative Logic. Int. J. Theor.
Phys. 21, 219-253 (1982)  \\\relax $\left [48\right ]$ Deutsch, D.: Quantum theory, the Church Turing principle and the
universal quantum computer. Proc. Roy. Soc. A 400, 97-117 (1985)  \\\relax $\left [49\right ]$ Aharonov, Y., Albert, D.Z. and Vaidman, L.: How the result of a
measurement of a component of the spin of a spin-1/2 particle can turn out to be 100. Phys. Rev. Lett. 60, 1351 (1988)  \\\relax $\left [50\right ]$ Elitzur, A.C., Cohen, E., Okamoto, R. and Takeuchi, S.: Nonlocal
position changes of a photon revealed by quantum routers. Sci. Rep. 8, 7730 (2018)  \\\relax $\left [51\right ]$ Cohen, E. and Pollak, E.: Determination of weak values of quantum
operators using only strong measurements. Phys. Rev. A 98, 042112 (2018)  \\\relax $\left [52\right ]$ Bennett, C. H. and Brassard, G.: Quantum cryptography: Public key
distribution and coin tossing. In Proceedings of IEEE International Conference on Computers, Systems and Signal Processing, vol. 175, page 8. New York (1984)
\\\relax $\left [53\right ]$ Ekert, A. K.: Quantum cryptography based on Bell's theorem. Phys.
Rev. Lett. 67, 661-663 (1991) 
\end{document}